\documentclass[a4paper,twoside]{article}
\newcommand{\affil}[1]{$^{\rm #1}$}
\usepackage[authoryear]{natbib}
\bibpunct{(}{)}{;}{a}{}{,}
\usepackage{graphicx}
\date{} %Please leave the date blank
\title{\large\bf\flushleft Initial Test of a Bayesian Approach to Solar 
Flare Prediction}
\author{\parbox{\textwidth}{\flushleft
\vspace{-0.5cm}
{\it M.S. Wheatland\affil{A,B}}\\
\vspace{0.4cm}
{\small \affil{A}\,School of Physics, University of Sydney, NSW 2006}\\
{\small \affil{B}\,Email: m.wheatland@physics.usyd.edu.au}}}
\begin{document}
\twocolumn[
\begin{changemargin}{.8cm}{.5cm}
\begin{minipage}{.9\textwidth}
\vspace{-1cm}
\maketitle
%
%
%%%%%%%%%%%%%     ABSTRACT    %%%%%%%%%%%%%
%Abstract of no more than 200 words here.
\small{\bf Abstract:} A test of a new Bayesian approach 
to solar flare prediction (Wheatland 2004a) is presented. The 
approach uses the past history of flaring together with phenomenological 
rules of flare statistics to make a prediction for the probability of 
occurrence of a large flare within an interval of time, or to refine 
an initial prediction (which may incorporate other information). 
The test of the method is based on data from the Geostationary 
Observational Environmental Satellites (GOES), and involves whole-Sun
prediction of soft X-ray flares for 1976-2003. The results show that 
the method somewhat over-predicts the probability of all events above a 
moderate size, but performs well in predicting large events.

%%%%%%%%%%%%%     KEYWORDS    %%%%%%%%%%%%%
\medskip{\bf Keywords:} methods: statistical --- Sun: activity --- Sun:
flares --- Sun: X-rays   
% Please write all keywords in lower case. PASA uses the
% standard list of subject headings adopted by The Astrophysical Journal
% and available from http://www.journals.uchicago.edu/ApJ/keywords_text.html.
% Keywords are separated by em-dashes, i.e. ---

%%%%%%%%DO NOT EDIT%%%%%%%%%%%%
\medskip
\medskip
\end{minipage}
\end{changemargin}
]
\small
%%%%%%%%EDIT FROM HERE%%%%%%%%%%%%

\section{Introduction}
%Please see the PASA Style Guide for help with correct layout for your manuscript.
%Examples of tables and figures are given below.

The space weather effects of large solar flares motivate flare
prediction. For example, the soft X-ray flux due to large flares causes
increased ionisation of the upper atmosphere, 
which can interfere with high frequency radio communication. There is
considerable interest in knowing when such short-wave fadeouts are 
likely to occur, and Australia's Ionospheric Prediction 
Service (IPS) issues predictions on this basis.\footnote{see
http://www.ips.gov.au} Other agencies issuing flare predictions
include the US National Oceanic and Atmospheric Administration 
(NOAA)\footnote{see http://www.sec.noaa.gov/ftpdir/latest/daypre.txt}
and NASA\footnote{see http://beauty.nascom.nasa.gov/arm/latest/}.

Existing methods of prediction are probabilistic, and
rely e.g.\ on the classification of physical characteristics of active
regions, and historical rates of flaring for regions with a given
classification (McIntosh 1990; Bornmann \& Shaw 1994). One weakness of
classification based approaches is that regions with a given classification 
may exhibit a wide range of flaring rates.
The method of McIntosh (1990) also considers other information including 
the number of large flares already produced by an active region (the 
tendency of an active region which has produced large events to 
subsequently produce large events is called persistence), but this is 
done in an ad hoc way. No consideration is given to the important 
information in the number of small events already observed.  

A new approach to flare prediction (Wheatland 2004a) exploits the 
history of observed flaring together with simple phenomenological rules 
of flare statistics to make a prediction, or to refine an existing 
prediction. The basic method is as follows. It is well known that the size
distribution of flares (e.g.\ the distribution of peak soft X-ray flux)
follows a power law (e.g.\ Crosby, Aschwanden \& Dennis 1993):
\begin{equation}\label{eq:fdist}
N(S)=\lambda_1(\gamma-1)S_1^{\gamma-1}S^{-\gamma},
\end{equation}
where $N(S)$ is the number of events per unit size $S$ and per unit
time, $\lambda_1$ is the total rate of events above size $S_1$, and
$\gamma$ is the power-law index. Suppose we are interested in the
probability of a large event ($S\geq S_2$) occurring in a time
$\Delta T$. The expected rate of events above $S_2$ is, according
to Equation~(\ref{eq:fdist})
\begin{equation}\label{eq:rate_big}
\lambda_2=\lambda_1
  \left( \frac{S_1}{S_2}\right)^{\gamma-1}.
\end{equation}
Flare occurrence may be described on short timescales as a Poisson 
process in time (e.g.\ Moon et al.\ 2001), and on longer timescales as 
a time-dependent Poisson process (e.g.\ Wheatland 2001). According to 
Poisson statistics, the probability of at least one large event in 
time $\Delta T$ is
\begin{equation}\label{eq:prob_big}
\epsilon = 1-\exp(-\lambda_2 \Delta T).
\end{equation}
To apply these formulae it is necessary to estimate $\lambda_1$ and
$\gamma$ from data, and hence to estimate $\epsilon$.
We adopt a Bayesian approach, in which `estimating' a parameter means
calculating a posterior probability distribution for the parameter, 
given available data and any prior information (e.g.\ Jaynes 2003).
We assume that a a sequence of $M$
events with sizes $s_1,s_2,...,s_M$ (all larger than $S_1$) have been
observed to occur at times $t_1< t_2< ...< t_M$ respectively.
The power-law index $\gamma$
may be approximated by the maximum likelihood value (Bai 1993)
\begin{equation}\label{eq:gam_ML}
\gamma^{\ast}=\frac{M}{\ln \pi}+1, \quad
{\rm where} \quad
\pi=\prod_{i=1}^M\frac{s_i}{S_1}.
\end{equation}
To estimate the rate we adopt a piecewise constant Poisson model. Hence
we need to identify the most recent interval $T^{\prime}$ during which
the rate is constant, and we assume that $M^{\prime}\leq M$
events occurred during that time. One approach to the determination of 
the interval $T^{\prime}$ is to use the `Bayesian blocks' procedure
(Scargle 1998), which is discussed in more detail below.
Based on $M^{\prime}$, $T^{\prime}$
and $\gamma^{\ast}$, the posterior probability distribution for
$\epsilon$ is (Wheatland 2004a)
\begin{eqnarray}\label{eq:pbig_simp}
P(\epsilon ) &=& 
  C\left[-\ln (1-\epsilon ) \right]^{M^{\prime}}
  (1-\epsilon )^{\left( T^{\prime}/\Delta T\right)
  \left(S_2/S_1\right)^{\gamma^{\ast}-1}-1} \nonumber \\
  &\times &
  \Lambda \left[-\frac{\ln (1-\epsilon )}{\Delta T}
  \left(\frac{S_2}{S_1}\right)^{\gamma^{\ast}-1} \right],
\end{eqnarray}
where $\Lambda (\lambda_1)$ is the prior distribution for $\lambda_1$,
i.e.\ the distribution we would assign to $\lambda_1$ in the absence
of any data, and $C$ is the normalisation constant, determined by the
requirement $\int_{0}^{1}P(\epsilon)d\epsilon=1$. The mean of 
$P(\epsilon )$ provides the estimate of the probability of at least one 
large event within time $\Delta T$, and the standard deviation of 
the distribution provides an estimate of the associated uncertainty.

\section{The test}

As a basic test of the new approach to prediction,
Equation~(\ref{eq:pbig_simp}) was applied to the NOAA Solar Event Lists
of X-ray flares observed by the Geostationary Observational Environmental 
Satellites (GOES) for 1975-2003.\footnote{available from
ftp://ftp.ngdc.noaa.gov} For each day, whole-Sun flare prediction was
performed for the next day ($\Delta T=1\,{\rm day}$). 

The relevant measure of size, $S$, of the GOES events is the peak 
soft X-ray flux in the 1-8\AA ~band. The choice of threshold 
$S_1=4\times 10^{-6}\,{\rm W}\,{\rm m}^{-2}$ 
was made, based on inspection of the distribution of peak soft X-ray 
flux for the entire dataset. Figure~1 shows the distribution, plotted in
differential form (upper panel) and cumulative form (lower panel).
The threshold size $S_1$ is indicated by the vertical solid line. 
Events above this size are observed to be distributed 
approximately as a power law, and the thick solid lines in each panel 
indicate the power-law model, with the maximum likelihood power-law index
$\gamma^{\ast}\approx 2.15\pm 0.01$. For peak fluxes below the threshold 
there is departure from power-law behaviour due to problems with
event selection against the time-varying soft X-ray background.

\begin{figure}[h]
\begin{center}
\includegraphics[scale=0.5, angle=0]{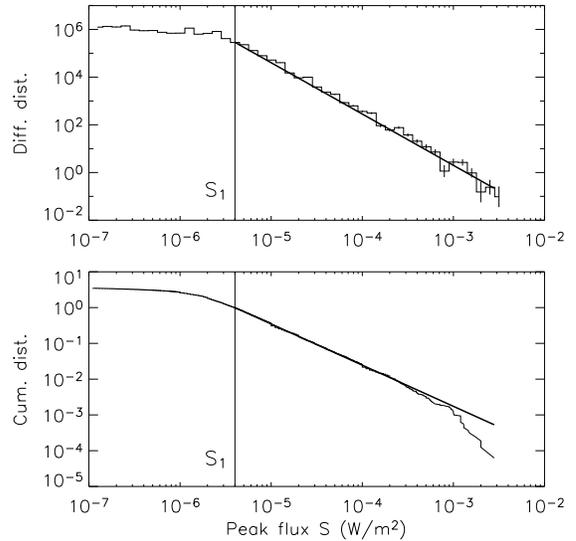}
\caption{Upper panel: differential distribution of peak flux for GOES 
events 1976-2003 (histogram), and the power-law model distribution 
(thick line). Lower panel: cumulative distribution for all events 
(joined points), and the model (thick line). In both panels the
threshold $S_1$ for power-law bevaviour is shown by the vertical 
line.}\label{f1}
\end{center}
\end{figure}

Predictions were made for each day for events above size
$S_2=10^{-5}\,{\rm W}\,{\rm m}^{-2}$ (`M class' and larger flares)
and for events above size $S_2=10^{-4}\,{\rm W}\,{\rm m}^{-2}$ (`X class' 
flares) using Equation~(\ref{eq:pbig_simp}). 
The corresponding prediction probabilities for a given day are 
labelled $\epsilon_{\rm M}$ and $\epsilon_{\rm X}$ 
respectively. It should
be noted that these values are not independent, since X class flares
are a subset of events above M class. 

The predictions used data within a window of time spanning
one year prior to each day. For each day, Equation~(\ref{eq:gam_ML})
was applied to the one-year window of data prior to the day to
determine $\gamma^{\ast}$. Then the Bayesian blocks procedure
was applied to the same data to determine a
decomposition into a piecewise-constant Poisson process. This
procedure returns a sequence of times
$t_{B0}< t_{B1}<...<t_{BK}$ at which the rate is determined
to change (where $t_{B0}$ and $t_{BK}$ are the start- and end-time of
the window), and a corresponding sequence
$\lambda_{B1},\lambda_{B2},...,\lambda_{BK}$
of rates. The last Bayesian block was used to
determine $T^{\prime}$ and $M^{\prime}$: viz.\
$T^{\prime}=t_{BK}-t_{B(K-1)}$ and
$M^{\prime}=\lambda_{BK}T^{\prime}$.

The data in every Bayesian block
but the last was used to construct the prior $\Lambda (\lambda_1)$.
A model form $\Lambda (\lambda_1)=a\exp (-b\lambda_1^c)$ was chosen for
the prior, and the parameters $a$, $b$, and $c$ were determined by
requiring that the first three moments of this model distribution match
the first three moments of the data, estimated from the Bayesian blocks
decomposition. Specifically we required that the model distribution was
normalised, and that it had a mean rate and mean square rate equal to
the corresponding estimates from the Bayesian blocks. 
These three conditions uniquely determined values of $a$, $b$, and $c$. 

\section{Results}

Figure~2 illustrates the results of the test on a year by year basis
for 1976-2003 (the results for 1975 are omitted because the predictions
are made using less than a year of previous data).
The upper panel shows the predictions for M class (and larger) flares. 
The histograms represent the observed number of days on which there was 
at least one M class flare (or larger). The diamonds represent the sum
of the $\epsilon_{\rm M}$ values for all days within the given year, which is 
the predicted number of days on which there should be at least one event 
of M class or larger. The lower panel shows the same display, but for 
events of X class. Uncertainties are shown for the predicted values,
based on summation of the individual prediction uncertainties in
quadrature.

The upper panel of Figure~2 indicates that the values of 
$\epsilon_{\rm M}$ are systematically too large. More quantitatively, we 
find that the average value of $\epsilon_{\rm M}$ over all days (1976-2003) 
is 0.320, whereas the observed value (the fraction of days on which there 
was at least one M class event) is 0.264. 
The lower panel of Figure~2 indicates that the method has done quite well
in predicting X class events. In fact the average value of 
$\epsilon_{\rm X}$ for 1976-2003 is 0.040, whereas the observed 
value is 0.036.

\begin{figure}[h]
\begin{center}
\includegraphics[scale=0.5, angle=0]{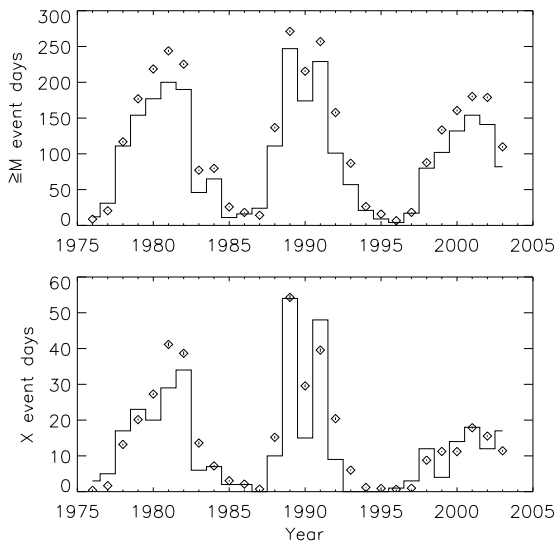}
\caption{Upper panel: observed/predicted numbers of M (or larger) event days 
(histogram/diamonds). Lower panel: the same but for X event days.}
\label{f2}
\end{center}
\end{figure}

Figure~3 gives a more detailed display of the results of the test
for all years (1976-2003) in the form of a pair of `reliability plots', 
in which the horizontal axis shows the forecast probabilities 
$\epsilon_{\rm M}$ or $\epsilon_{\rm X}$ for 
each day (in bins of 0.05), and the vertical axis shows the true
probability for flaring on the given days, estimated from the
observed number of event days. This is the Bayesian estimate
assuming binomial statistics and a uniform prior: if there are $R$ days
with at least one event out of a total of $S$ days, the estimate for
the probability
is $p=(R+1)/(S+2)$ and the corresponding error is $[p(1-p)/(S+3)]^{1/2}$
(e.g.\ Jaynes 2003, p.\ 165). Perfect prediction corresponds to the
solid 45 degree line on the plot. 
The upper panel of Figure~3 is the reliability plot for M (and larger) 
event prediction, and the lower panel is the reliability plot for X event 
prediction. The upper panel confirms that the predictions for M class 
events are systematically too large, although it shows that the effect is 
only associated with days on which $\epsilon_{\rm M}$ is larger 
than about 0.25. The lower panel indicates that the predictions for X class 
flares are quite good for all values of $\epsilon_{\rm X}$, although the 
method is conservative, in that it does not assign values larger than about 
0.5.

\begin{figure}[h]
\begin{center}
\includegraphics[scale=0.5, angle=0]{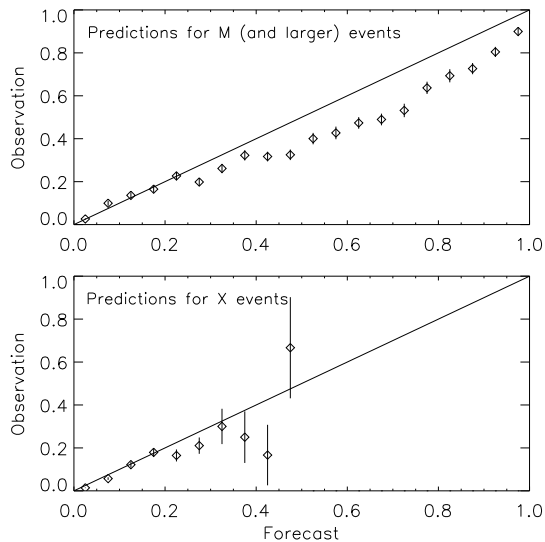}
\caption{Upper panel: reliability plot for prediction of M events and 
above. Lower panel: the same but for X events.}\label{f3}
\end{center}
\end{figure}

It is interesting to compare these results with predictions made by the US 
National Oceanic and Atmospheric Administration. Statistics are available
for NOAA predictions for 1987-2002\footnote{See
http://www.noaa.sec.gov/verification/}. The NOAA results indicate a very
serious over-prediction of X class events. During 1987-2002 the NOAA 
predictions imply that there should be 372.5 X event days, when in 
fact there are 200 such days. 
The present method predicts 233.8 X event days for
1987-2002, which is a considerable improvement over the NOAA result. For M
(and larger) events, the NOAA results show over-prediction similar to the 
results obtained with the present method. A more detailed comparison
with NOAA predictions will be presented in a future paper (Wheatland 
2004b). 

\section{Discussion}

A Bayesian approach to flare prediction (Wheatland 2004a) has been tested 
for whole-Sun prediction of GOES soft X-ray events, based on the NOAA
Solar Event Lists for 1975-2003. 
The method is found to over-predict events of M class and above, but 
performs quite well for prediction of X class events.

There are several possible reasons for the over-prediction of events
above M class. One possibility is that the method is systematically late 
in detecting the decline in rate associated with the decay of a large 
active region, or the rotation of a large active region off the disk. 
The method uses the Bayesian blocks procedure to detect rate changes,
and is always trying to `catch up' with what the Sun is doing. A
specific limitation of the Bayesian blocks method is that it must have
at least one event in a block, so that the rate is never identically
zero. This limitation leads to overestimation of the rate at times of
very low activity. It should also be noted that the existing Bayesian 
blocks procedure is not guaranteed to find an optimal decomposition --- 
in future this procedure will be replaced by an optimal algorithm recently 
devised by Scargle (Scargle, private communication, 2003). 
Another possibility is that there is a bias in 
each individual prediction which becomes apparent in the analysis of 
a large number of predictions. Such a bias will become less serious if 
each individual prediction is more accurate. The fractional error in 
each prediction goes as $(M^{\prime})^{-1/2}$ (Wheatland 2004a),
where $M^{\prime}$ is the number of events associated with the 
estimation of the rate above size $S_1$. This error becomes smaller with
increasing $M^{\prime}$, i.e.\ if a smaller $S_1$ can be chosen. 
It should be noted that the GOES event lists are a relatively poor choice 
for the present purpose because the time varying soft X-ray background means 
that a relatively large $S_1$ must be used (see Figure~1). 

The GOES event lists also have other shortcomings as a basis for 
prediction from event statistics. Besides the departure from a power
law at small sizes, it is likely the lists are incomplete above
the nominal threshold $S_1$, e.g.\ due to the difficulty of
distinguishing two flares occurring close together in time 
(Wheatland 2001). This is unlikely to be important for the test described 
here, provided that the size distribution obeys a power 
law above the threshold, since both prediction and validation of the 
prediction rely on the same (possibly incomplete) lists. 
Another problem with the GOES event lists is that the GOES peak fluxes 
are not background subtracted, so that the intrinsic flux due to an M 
class event at a time of low activity, when the background is low, is 
larger than the intrinsic flux due to an M class event at a time of high 
activity, when the background is high.  However, again this is not 
particularly important to the present method provided that the size 
distribution for observed events is not distorted from a power-law form 
by this effect. The use of a previous year of data to make a prediction 
allows the possibility of incorporating variation in the power-law index 
with the solar cycle (e.g.\ due to this effect) but in fact we find no 
evidence of such a variation. In subsequent work the method will be applied 
to more accurate event catalogs.

The present test is limited to whole-Sun prediction. In the future the 
method will also be applied to individual active regions, and other 
prior information on the rate (e.g.\ the McIntosh classification of the 
associated sunspots) will be incorporated. However we note that, even 
in this simple form, the method has out-performed NOAA predictions for 
X class events. 

Finally we note that automated predictions, based on this 
method, are now published on the web.\footnote{see
http://www.physics.usyd.edu.au/$\sim$wheat/prediction/prediction.html} 
Predictions are made each day using the latest NOAA solar event
lists. The web pages include a running score of how reliable the published 
predictions are, in the form of automatically updated reliability plots. 

\section*{Acknowledgments} %If needed
The author acknowledges the support of an ARC QEII Fellowship. The data
used here is from the Space Environment Center of the US National
Oceanic and Atmospheric Administration (NOAA).

%\end{multicols}

\end{document}